\documentclass[doublecol]{epl2}
\pdfoutput=1
\usepackage[T1]{fontenc}
\usepackage[latin9]{inputenc}
\usepackage{textcomp}
\usepackage{amsmath}
\usepackage{amssymb}
\usepackage{graphicx}
\usepackage{esint}

\title{Excitation of SPP's in graphene by a waveguide mode}

\author{D. C. Pedrelli\inst{1,2}, B. S. C.  Alexandre\inst{2,3} \and N. M. R. Peres\inst{2,3}}
\institute{
	\inst{1} Faculdade de Física, Universidade Federal do Pará - 66075-110, Belém,
	Pará, Brazil\\
	\inst{2} Departamento e Centro de Física, Universidade do Minho - 4710-057, Braga, Portugal\\
	\inst{3} International Iberian Nanotechnology Laboratory (INL) - 4715-330, Braga, Portugal
	
	{Optical Properties}	
	{Plasmons}	
	{ Graphene }	
	
}
\abstract{
We present a semi-analytical model that predicts the
excitation of surface-plasmon polaritons (SPP) on a graphene sheet located in front of a sub-wavelength slit drilled in thick metal screen.  We identify the signature of the SPP in the transmission, reflection, and absorption curves. Following previous literature on noble-metal plasmonics, we characterize the efficiency of excitation of SPP's in graphene computing a spatial probability density. This quantity shows the presence of plasmonics resonances dispersing with the Fermi energy, $E_F$, as $\sqrt{E_F}$ an unambiguous signature of graphene plasmons.
}

\begin{document}
	
\maketitle

\section{Introduction}

Surface-plasmon polaritons (SPPs) are localized  in space electromagnetic fields oscillating at the interface between a metal and a dielectric. Photon confinement and its application to nano-optics  is only one example where SPPs have been explored to the creation of new sub-wavelength technologies. This  has caused an increasing interest in studying this type of electromagnetic fields (see \cite{Lalanne2009-SSP} and references therein).

However, due to the mismatch between photon momentum in the free space and the SPP momentum, shinning  light directly on a metallic-dielectric interface will not generate SPPs  \cite{barnes2003surface}. Therefore, a large number of structures have been investigated in order to understand how SPPs are produced. The central question is about the most effective way to achieve their excitation. Traditional forms of exciting SPPs encompasses the total attenuate reflection scheme, the use of gratings, and shinning electromagnetic radiation at wedges and slits. Among these structures, metal slits \cite{takakura2001optical,popov2007single,chandran2012metal}, gratings \cite{todorov2007modal,lee2005coupling}, and indentations allow the excitation of localized surface-plasmons \cite{lopez2005scattering}.

Theoretically, the excitation of SPPs requires that the metallic structures are treated beyond the perfect metal approximation. This can be done in different ways, either modeling the metal by sophisticated Drude and Lorentz models, or assuming the surface impedance boundary condition (SIBC) \cite{lopez2005scattering,lee2005coupling}, the latter having the advantage that does not require solving Maxwell's equations in the metal to obtain the fields in the space surrounding the metallic nano-structure.

The scattering problem of such structures has been studied long before SPPs became an active field in both nano-optics and  surface science \cite{bethe1944theory,bouwkamp1954diffraction,ritchie1957plasma,roberts1987electromagnetic}. In this context,   Kang,  Eom, and Park \cite{park1993analytic,kang1993tm} were, to our best knowledge, the first to describe, in a closed theoretical form, the behavior  of light propagation through a slit aperture on a thick metallic screen. On the other hand, most of the former works have developed either semi-analytical methods or powerful  computational tools \cite{wannemacher2001plasmon,chang2005surface,xie2004transmission,lalanne2005surface} in order to make quantitative predictions of how light propagates through apertures. However, they did not consider the SPP generation at the input or output sides of the slit \cite{lalanne2006_approximate}.

The interest in SPPs received a strong boost after the pioneering work of  Ebbesen \emph{et al.} \cite{ebbesen1998extraordinary}, since the excitation of SPPs in the slits of the metallic screen was at the root  of the interpretation of the so  called extraordinary optical transmission (EOT).  This    unexpected phenomenon is characterized by sharp peaks on the transmission spectrum  when light passes through sub-wavelength indentations on a thin metal film. However, Gay \emph{et al.}\cite{gay2006optical}\emph{} have put forward an alternative  explanation of  EOT.  In their interpretation, EOT exists due to a new type of surface wave, called by those authors  "composite diffracted evanescent waves'' (CDEW). Their interpretation, was argued, should have a  better agreement with their experimental results than the SPP model.

An important contribution to decide whether or not SPPs are decisive to describe EOTs has been given in a paper by Lallane \emph{et al.} \cite{lalanne2006interaction}. In this work, the authors concluded that Gay \emph{et al.} results are due to some possible experimental error, such as impurities on the film they have used.

Considering one-dimensional lamellar metallic gratings with sub-wavelength slit apertures, Lallane \emph{et al.} \cite{lalanne2000one} have developed a simple semi-analytical model for determining  the transmission through a grating, relying on the fact that metallic films behave as monomode waveguides.

In two other  papers \cite{Lalanne2005-PRL,lalanne2006_approximate}, Lallane \emph{et al.} built a semi-analytical model that predicts the creation of SPPs due to a sub-wavelength slit in a thick metal screen. Their results were shown to be in excellent agreement with computational and experimental results, even for noble metals with low conductivity. The work is based on a two stage scattering mechanism where  they considered the diffraction problem in the vicinity of the aperture, followed by the launching of SPPs on the metallic-dielectric interface.

SPP's in graphene have recently created a new field of research in nanophotonics, generating an intense interaction between experimental results,  technological applications, and analytical and numerical modeling \cite{Review_Asgar}. Graphene surface-plasmon polaritons provide some advantages over noble metal plasmonics (in the infrared range), such as strong spatial confinement and long propagation lengths, with the additional benefit of  being electrically and chemically tunable \cite{liu2011graphene,ding2015effective,phare2015graphene,goykhman2016chip}.

In this context, there is the need for discussing the excitation of surface-plasmons on a graphene sheet covering a slit on a thick metal screen. In the present work, we shall consider a TM polarized field propagating in a waveguide and emerging from an aperture covered by graphene. We assume that the field propagating in the wave-guide is  the fundamental mode, an approximation also considered in Refs. \cite{Lalanne2005-PRL,lalanne2006_approximate}. Quantitatively, we compute how efficiently will the power be transmitted from the outgoing mode to the graphene plasmons. In other words, we must obtain the transmission, reflection, and absorption of the electromagnetic radiation impinging on the graphene sheet, as well as the spatial probability density of exciting surface-plasmons.

\section{Modes in the cavity and field amplitudes: emission of a wave at the
aperture of a slit\label{sec:Modes-in-the}}

\begin{figure}[h]
\centering{}\includegraphics[width=7cm]{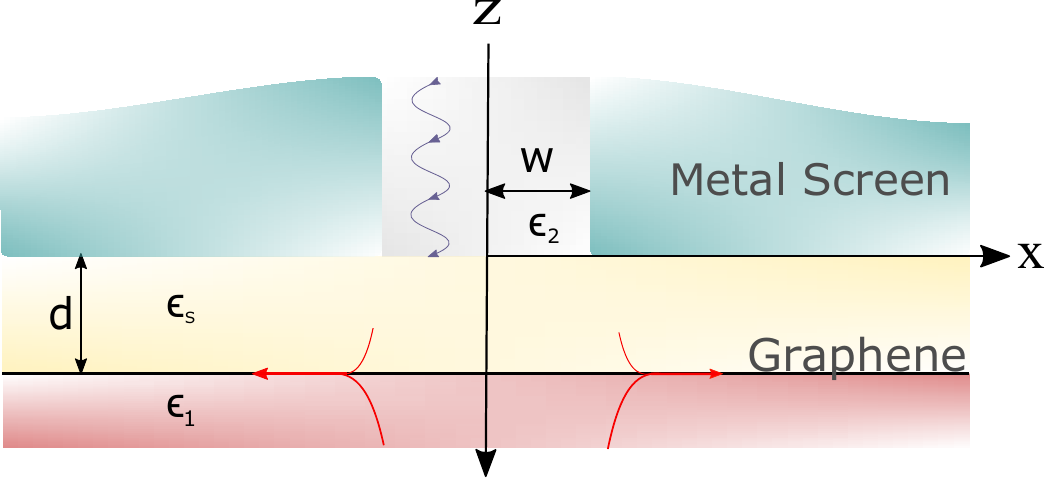}\caption{The geometry studied in this work. A TM polarized radiation,  emerging for an aperture (of width $2w$) on a perfect metal screen, passes through an intermediary media and impinges on graphene, thus  creating a SPP at the graphene sheet, which is located at a distance $d$ from the metal slit. \label{Fig-boneco}}
\end{figure}

The geometry of the problem we are dealing with in this paper is represented in Fig. \ref{Fig-boneco}. In region $2$ ($z<0$, $-w<x<w$), an incident TM polarized radiation propagates through a slit aperture on a semi-infinite perfect metal screen, thus producing a reflected and a transmitted component. The dielectric constant of the wave-guide is $\epsilon_2$, the dielectric constant between graphene the metallic screen is $\epsilon_s$, and, finally, the dielectric constant bellow graphene is $\epsilon_1$. After the radiation emerges from the slit, it  propagates until reaching the graphene sheet, where reflected and transmitted waves are created. Finally, in region $1$ the fields propagate freely. In what follows, we assume a time dependence of the fields of the form $e^{-i\omega t}$, which is  suppressed in all the  equations below.

\subsection{Waveguide modes\label{subsec:Wave-guide-modes}}

As already mentioned,  the slit acts as a waveguide and the impinging radiation is propagating along it. Assuming that the metallic screen is a perfect electric conductor, the tangential component of the electric field must vanish at the slit walls, and, since we are considering a TM polarization, the electric field will have only the $x$ and $z$ components. On the other hand,  the magnetic field has only a finite $y$ component. The modes propagating on a semi-infinite slit have already been calculated by Park \emph{et al.} \cite{TE-polarization}.
We can separate the magnetic field into its fundamental part plus
the remaining modes, thus obtaining
\begin{equation}
H_{y}^{(2)}=b_{0}e^{i\zeta_{0}z}\Psi_{0}(x)+\sum_{m=1}^{\infty}b_{m}e^{i\zeta_{m}z}\Psi_{m}(x),
\end{equation}
where $\zeta_{0}=\sqrt{\epsilon_{2}}\omega/c$, $\Psi_{0}(x)=\theta(w-\vert x\vert)$,  $w$ is half the slit size, $\zeta_{m}=\sqrt{\epsilon_{2}\omega^{2}/c^{2}-a_{m}^{2}}$,
$a_{m}=m\pi/(2w)$, and $b_{m}$ is the mode amplitude. The electric
field components follow from Maxwell's equations and read  $\Psi_{m}(x)=\cos[a_{m}(x-w)]$. We will consider only the fundamental mode as the impinging one, which is the core approximation of the method proposed here and  also considered in the past \cite{lalanne2006_approximate,Lalanne2005-PRL}. The constant $b_{0}$ is fixed by the power per unit length of the incoming mode:
\begin{equation}
{\cal P}_{z}^{(2)}=\frac{1}{2}\int_{-w}^{w}dx\Re[\mathbf{E}\times\mathbf{H}^{\ast}]\cdot\mathbf{e}_{z}=w\vert b_{0}\vert^{2}\frac{\mu_{0}c}{\sqrt{\epsilon_{2}}},\label{power-2}
\end{equation}
where, from here on, we  consider $|b_{0}|=1$. Within the introduced
approximation, and considering the reflected part of the wave, the
magnetic field at region $2$ turns out to be
\begin{equation}
H_{y}^{(2)}(x,z)=(e^{i\zeta_{0}z}+r_{0}e^{-i\zeta_{0}z})\Psi_{0}(x),\label{mag-2}
\end{equation}
where $r_{0}$ is the reflection coefficient. The electric field,
then, becomes
\begin{align}
E_{x}^{(2)}(x,z) & =\frac{\mu_{0}c^{2}}{\epsilon_{2}\omega}\zeta_{0}(e^{i\zeta_{0}z}-r_{0}e^{-i\zeta_{0}z})\Psi_{0}(x),\label{elec-x-2}
\end{align}
\begin{equation}
E_{z}^{(2)}(x,z)=0.\label{elec-z-2}
\end{equation}

Now that we have defined the waveguide modes, lets determine the scattered modes, which propagate on regions $S$ and $1$.

\subsection{Scattered modes\label{subsec:Scattered-modes}}

Differently to the waveguide modes, which are finite and have a discrete
representation, the scattered modes are continuous and represented
by a Fourier transform. In region $S$, the one between the metal
and the graphene sheet, we write the fields as
\begin{equation}
H_{S,y}=\int_{-\infty}^{\infty}dk\,e^{ikx}(A_{k}e^{i\zeta_{S,k}z}+B_{k}e^{-i\zeta_{S,k}z}),\label{mag-s}
\end{equation}
\begin{equation}
E_{S,x}=\frac{\mu_{0}c^{2}}{\epsilon_{S}\omega}\int_{-\infty}^{\infty}dk\,e^{ikx}\zeta_{S,k}(A_{k}e^{i\zeta_{S,k}z}-B_{k}e^{-i\zeta_{S,k}z}),\label{eq:ESx}
\end{equation}
\begin{equation}
E_{S,z}=-\frac{\mu_{0}c^{2}}{\epsilon_{S}\omega}\int_{-\infty}^{\infty}dk\,ke^{ikx}(A_{k}e^{i\zeta_{S,k}z}+B_{k}e^{-i\zeta_{S,k}z}),\label{eq:ESz}
\end{equation}
where $\zeta_{S,k}=\sqrt{\epsilon_{S}\omega^{2}/c^{2}-k^{2}}$, and
$A_{k}$ and $B_{k}$ are the incident and reflected amplitudes, respectively.
For region $1$, below the graphene sheet, we have
\begin{equation}
H_{1,y}(x,z)=\int_{-\infty}^{\infty}dk\,t_{k}e^{ikx}e^{i\zeta_{1,k}z},\label{mag-1}
\end{equation}
\begin{equation}
E_{1,x}(x,z)=\frac{\mu_{0}c^{2}}{\omega\epsilon_{1}}\int_{-\infty}^{\infty}dk\,t_{k}\zeta_{1,k}e^{ikx}e^{i\zeta_{1,k}z},\label{eq:E1x}
\end{equation}
\begin{equation}
E_{1,z}(x,z)=-\frac{\mu_{0}c^{2}}{\omega\epsilon_{1}}\int_{-\infty}^{\infty}dk\,kt_{k}e^{ikx}e^{i\zeta_{1,k}z},\label{eq:E1z}
\end{equation}
where $\zeta_{1,k}=\sqrt{\epsilon_{1}\omega^{2}/c^{2}-k^{2}}$ and
$t_{k}$ is the transmission amplitude. A straightforward result is
the propagating power along the $z-$direction, which is given by
\begin{align}
{\cal P}_{z}^{(1)} =\pi\frac{\mu_{0}c^{2}}{\epsilon_{1}\omega}\int_{-\sqrt{\epsilon_{1}}\omega/c}^{\sqrt{\epsilon_{1}}\omega/c}dk\,\zeta_{k}\vert t_{k}\vert^{2}.\label{power-1}
\end{align}

As we can see from the above equations, in order to fully determine the waveguide and scattered
modes is necessary to obtain the explicit expressions for the amplitudes $r_{0}$, $A_{k}$,
$B_{k}$, and $t_{k}$. In the next section, we use the boundary conditions for determining the previous scattering coefficients.

\subsection{The boundary conditions\label{subsec:The-boundary-conditions}}

Since the screen is composed by perfect metal, the fields are zero for
$z<0$ and $x<-w$ or $x>w$. The boundary conditions
\begin{equation}
H_{y}^{(2)}(x,0)=H_{y}^{(S)}(x,0),\label{eq:bound-1}
\end{equation}
and
\begin{equation}
E_{x}^{(2)}(x,0)=E_{x}^{(S)}(x,0),\label{eq:bound-2}
\end{equation}
apply in the aperture at $z=0$.
Assuming that graphene is a perfect two-dimensional material, the
other two boundary conditions that  hold at $z=d$ are: \cite{gonccalves2016introduction}:
\begin{equation}
E_{S,x}(x,d)=E_{1,x}(x,d),\label{eq:bound-3}
\end{equation}
\begin{equation}
H_{S,y}(x,d)-H_{1,y}(x,d)=\sigma E_{1,x}(x,d),\label{eq:bound-4}
\end{equation}
which implies the continuity of the tangential component of the electric
field and the discontinuity of the tangential component of the magnetic
field through the graphene  interface, due to the finite conductivity $\sigma(\omega)$ of graphene. 
For modeling the conductivity, and for simplicity, we represent it by the Drude conductivity, $\sigma(\omega)$  \cite{gonccalves2016introduction}:
\begin{equation}
\sigma(\omega,E_F)=\frac{4\epsilon_{0}c\alpha E_{F}}{\Gamma-i\hbar\omega},\label{Conductivity}
\end{equation}
where $\alpha$ is the fine structure constant
$
\alpha=e^{2}/(4\pi\epsilon_{0}\hbar c)
$, $E_F$ is the Fermi energy of graphene, and, hereafter, we will take $\Gamma=10^{-3}\,$eV. Note that this is a good approximation if we work in the infrared part of the electromagnetic spectrum and have graphene not too close to the metal slit, since, otherwise, nonlocal effects woud become important.

Inserting the electric and magnetic fields for the three
regions in the boundary conditions  (\ref{eq:bound-1})
to (\ref{eq:bound-4}), we obtain
\begin{equation}
\left(1+r_{0}\right)=\int_{-\infty}^{\infty}dk\left(A_{k}+B_{k}\right){\rm sinc}(kw),\label{eq:r_0}
\end{equation}
\begin{equation}
\zeta_{2,0}\frac{\epsilon_{S}}{\epsilon_{2}}\left(1-r_{0}\right){\rm sinc}(kw)=\zeta_{S,k}\left(A_{k}-B_{k}\right)\frac{\pi}{w},
\end{equation}

\begin{equation}
\frac{\epsilon_{1}}{\epsilon_{S}}\zeta_{S,k}\left(A_{k}e^{i\zeta_{S,k}d}-B_{k}e^{-i\zeta_{S,k}d}\right)=t_{k}\zeta_{1,k}e^{i\zeta_{1,k}d},
\end{equation}

\begin{equation}
A_{k}e^{i\zeta_{S,k}d}+B_{k}e^{-i\zeta_{S,k}d}=\left(\sigma\frac{\mu_{0}c^{2}}{\omega\epsilon_{1}}\zeta_{1,k}+1\right)e^{i\zeta_{1,k}d}t_{k},
\end{equation}
which is a system of four equations whose solution can be easily obtained. Since we cannot 
perform the integral in Eq. (\ref{eq:r_0}),  we first  find the solutions
for $A_{k}$, $B_{k}$, and $t_{k}$ in terms of $r_{0}$:
\begin{align}
A_{k} & =\frac{1}{\Delta}(1-r_{0})w\epsilon_{S}{\rm sinc}(kw)\zeta_{2,0}\nonumber \\
 & \qquad\times\left[\epsilon_{1}\zeta_{S,k}+\zeta_{1,k}\left(\epsilon_{S}+\frac{\sigma}{\epsilon_{0}\omega}\zeta_{S,k}\right)\right],\label{sol:A_k}
\end{align}
\begin{align}
B_{k} & =\frac{1}{\Delta}e^{2i\zeta_{S,k}d}(1-r_{0})w\epsilon_{S}{\rm sinc}(kw)\zeta_{2,0}\nonumber \\
 & \qquad\times\left[\epsilon_{1}\zeta_{S,k}+\zeta_{1,k}\left(-\epsilon_{S}+\frac{\sigma}{\epsilon_{0}\omega}\zeta_{S,k}\right)\right],\label{sol:B_k}
\end{align}
\begin{equation}
t_{k}=\frac{2}{\Delta}e^{i(\zeta_{S,k}-\zeta_{1,k})d}(1-r_{0})w\epsilon_{1}\epsilon_{S}{\rm sinc}(kw)\zeta_{2,0}\zeta_{S,k},\label{sol:t_k}
\end{equation}
where
\begin{align}
\Delta & =\pi\epsilon_{2}\zeta_{S,k}\left\{ \left(1-e^{2i\zeta_{S,k}d}\right)\epsilon_{1}\zeta_{S,k}\right.\nonumber \\
 & \quad\left.+\zeta_{1,k}\left[\left(1+e^{2i\zeta_{S,k}d}\right)\epsilon_{S}+\left(1-e^{2i\zeta_{S,k}d}\right)\frac{\sigma}{\epsilon_{0}\omega}\zeta_{S,k}\right]\right\} .\label{sol:delta}
\end{align}
Now, replacing Eqs. (\ref{sol:A_k}) and (\ref{sol:B_k}) into Eq.
(\ref{eq:r_0}), we obtain
\begin{align}
r_{0} & =\frac{J-1}{1+J},\label{r0}
\end{align}
where
\begin{align}
J & =\int_{-\infty}^{\infty}dk\frac{{\rm sinc}^{2}(kw)}{\Delta}w\epsilon_{S}\zeta_{2,0}\left\{ (1+e^{2i\zeta_{S,k}d})\epsilon_{1}\zeta_{S,k}\right.\nonumber \\
 & \left.+\zeta_{1,k}\left[(1-e^{2i\zeta_{S,k}d})\epsilon_{S}+(1+e^{2i\zeta_{S,k}d})\frac{\sigma}{\epsilon_{0}\omega}\zeta_{S,k}\right]\right\} .\label{P-integral}
\end{align}

The above integral cannot  be solved analytically. 
Thus, the method of analysis of the scattering problem is analytically up to the calculation of the $J$.
Since we have found $A_{k}$, $B_{k}$,
$r_{0}$, and $t_{k}$, the electric and magnetic fields became completely
determined. The next step is to evaluate how efficiently the
power per unit area is transmitted from the aperture to graphene plasmons.
One method of evaluating this coupling is  computing the transmission, reflection, and absorption.

Finally, we note that although we have approximated the field in the slit by a single propagating mode, this approximation is valid since all the other modes of the slit are located at a much larger frequency and therefore they are all evanescent in nature. It would not be very difficult to include some of the low lying evanescent modes in the calculation. But this would imply a cumbersome (and unnecessary) expression for $\cal R$ and $\cal T$.

\subsection{Reflection, transmission and absorption\label{subsec:The-reflectance-and}}

\begin{figure}
\centering{}\includegraphics[width=7cm]{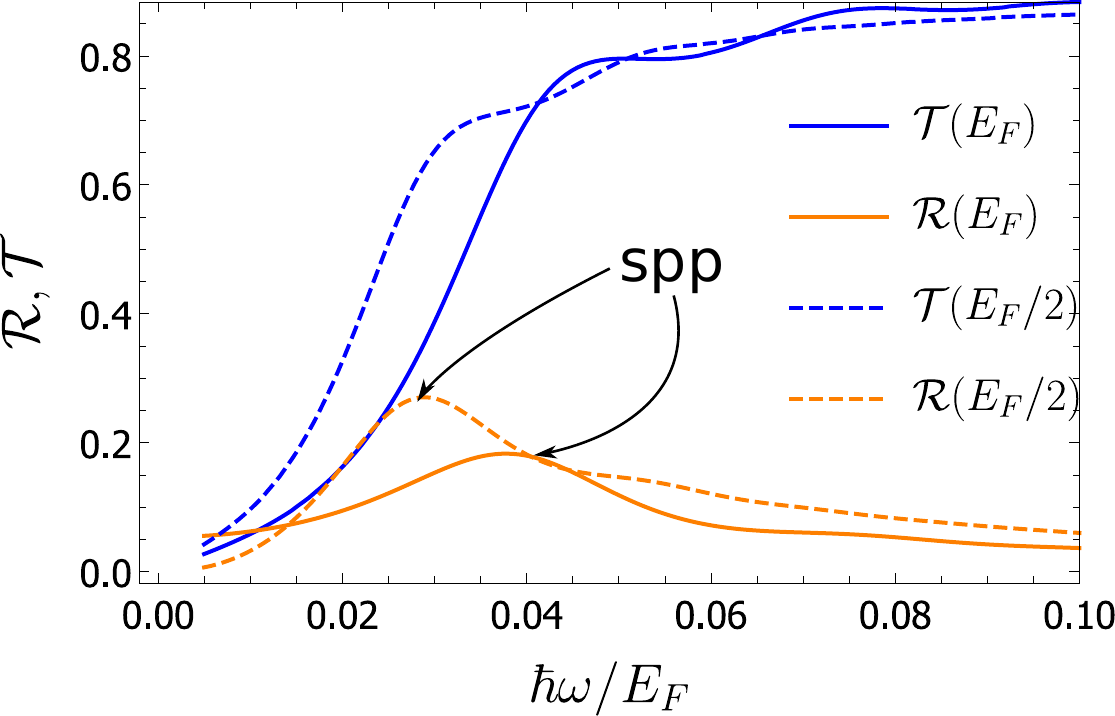}\caption{Reflection and transmission, for two values of Fermi energy, in
terms of the frequency per Fermi energy, with $E_{F}=1\,$eV,
$\epsilon_{1}=1$, $\epsilon_{S}=1.5$, $\epsilon_{2}=1$, $w=2\,\mu \rm{m}$, $\Gamma=10^{-3}\,$eV
and $d=0.3\,\mu \rm{m}$. The peaks at the reflection curves correspond
to the excitation of  SPPs at those frequencies, which are approximately $\omega/E_{F}\approx0.028$
for $\mathcal{R}(E_{F}/2)$ and $\hbar\omega/E_{F}\approx0.04$ for $\mathcal{R}(E_{F})$.
Note that the peak of the reflection disperses
with $E_F$. This is a plasmonic effect, where the 
reflection is enhanced by plasmonic assistance.
\label{fig:Reflectance-and-transmittance}}
\end{figure}

\begin{figure}
\centering{}\includegraphics[width=7cm]{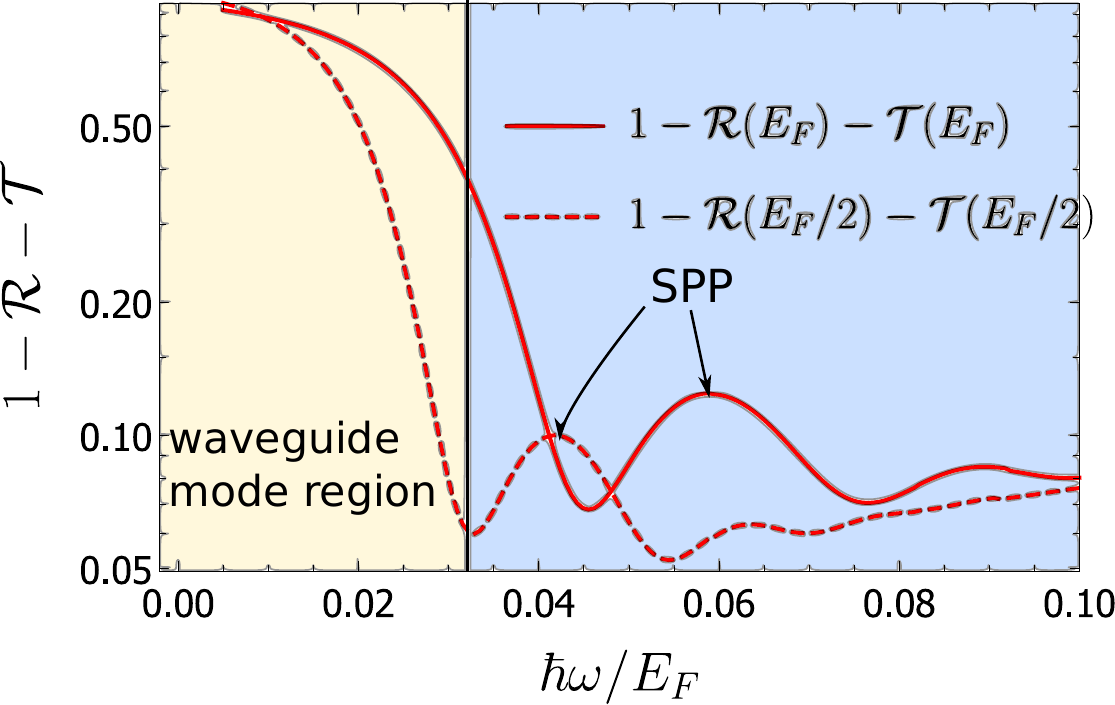}
\centering{}\includegraphics[width=7cm]{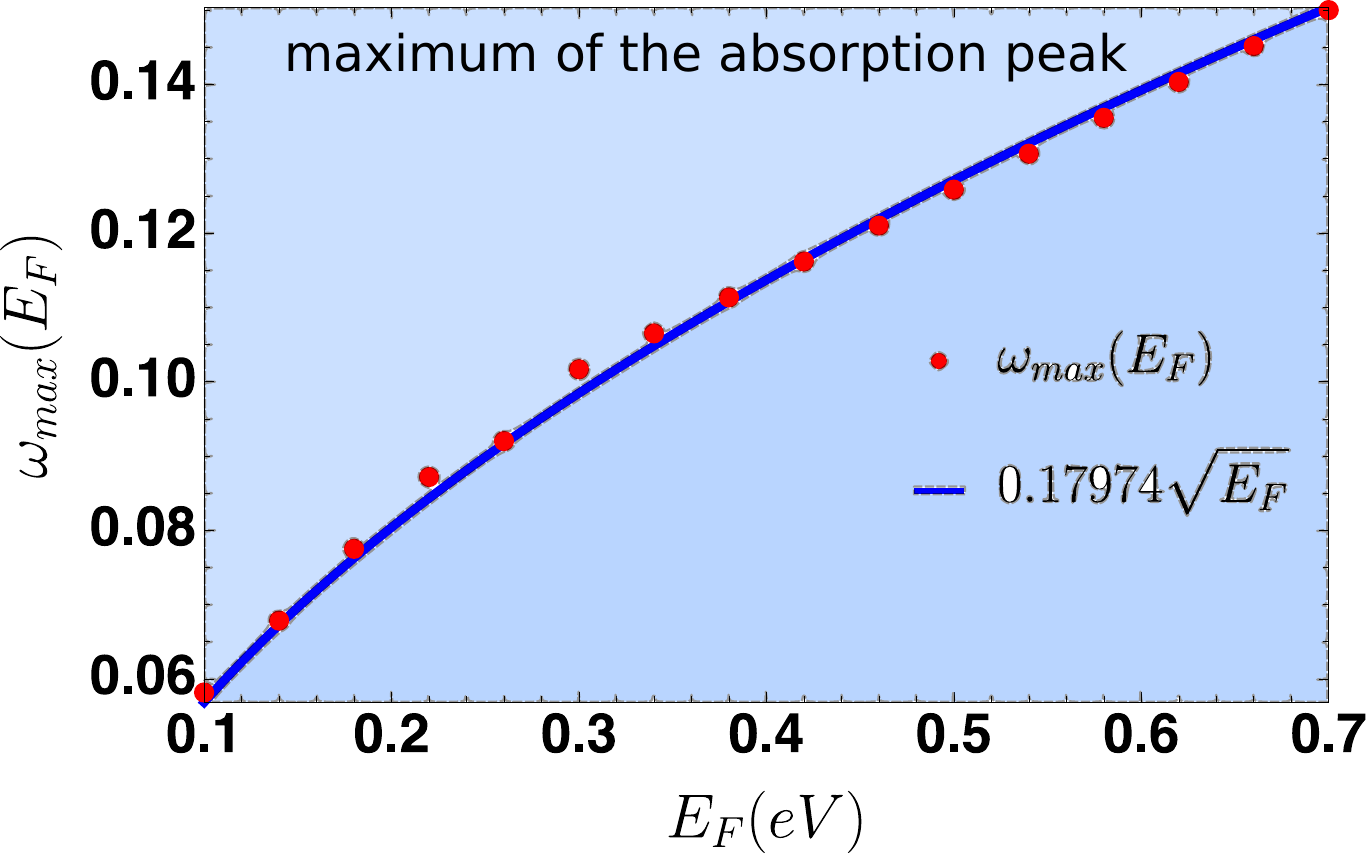}\caption{Top panel: Logarithmic plot of the absorption  for two values of Fermi energy (same as in Fig. \ref{fig:Reflectance-and-transmittance}), with  $E_{F}=1\,$eV,
$\epsilon_{1}=1$, $\epsilon_{S}=1.5$, $\epsilon_{2}=1$, $w=2\,\mu \rm{m}$, $\Gamma=10^{-3}\,$eV
and $d=0.3\,\mu \rm{m}$.
The figure can be roughly divided into two regions: a low energy one, where the excitation of waveguide modes between graphene and the metal dominate, and a high energy region, where we identify the excitation of surface plasmons in graphene.
Bottom panel: dispersion of the frequency at the maximum of the absorption curves, associated with the excitation of SPPs,
as function of the Fermi energy (for $w=1\,\mu \rm{m}$): there is a clear 
square-root dependence on the Fermi energy.
\label{fig:Absorbance}}
\end{figure}

The reflection and transmission are, respectively, defined as:
$
\mathcal{R}=|r_{0}|^{2}$
and
${\cal T}={\cal P}_{z}^{(1)}/{\cal P}_{z}^{(2)},
$
which, from Eqs.  (\ref{power-2}), (\ref{power-1}), and (\ref{r0}), gives
$
\mathcal{R}=\left|(1-J)/(1+J)\right|^{2},
$
\begin{equation}
\mathcal{T}=\frac{2\pi c\sqrt{\epsilon_{2}}}{w\epsilon_{1}\omega}\int_{0}^{\sqrt{\epsilon_{1}}\omega/c}dk\,\zeta_{1,k}|t_{k}|^{2}.
\end{equation}
In Fig. \ref{fig:Reflectance-and-transmittance}, we plot the curves
for  $\mathcal{R}$ and $\mathcal{T}$ in terms of $\hbar\omega/E_{F}$,  considering two different Fermi energies. That is,  we are computing $\mathcal{R}$ and $\mathcal{T}$ for different
values of the conductivity, $\sigma(\omega,E_{F}/2)$ and $\sigma(\omega,E_{F})$.
As an example, we have calculated the reflection and  transmission (see Fig. \ref{fig:Reflectance-and-transmittance}), and absorption (see Fig. \ref{fig:Absorbance}) 
curves for Fermi energy $E_{F}=1$ eV,
$d=0.3\,\mu \rm{m}$, and half of the slit aperture  $w=2\,\mu \rm{m}$. We can see that there are peaks in the reflection at
$\hbar\omega/E_{F}\approx0.028$ for $\mathcal{R}(E_{F}/2)$ (orange dashed
line), and $\hbar\omega/E_{F}\approx0.04$ for $\mathcal{R}(E_{F})$ (orange
solid line). These peaks are identified with the excitation of SPPs in graphene.
It is well known that $\omega_{SPP}\propto\sqrt{E_{F}}$, therefore, we must have
\begin{equation}
\frac{\omega_{SPP}(E_{F1})}{\omega_{SPP}(E_{F2})}=
\sqrt{\frac{E_{F1}}{E_{F2}}}.
\label{eq_ratio}
\end{equation}
This result implies that for peaks associated with $E_{F1}=2E_{F2}$,
Eq. (\ref{eq_ratio}) predicts a ratio of $\sqrt{2}\approx1.4$.
For the case of Fig. \ref{fig:Reflectance-and-transmittance},
we find from inspection of the curves $\omega_{SPP}(E_{F})/\omega_{SPP}(E_{F}/2)\approx1.4$,
which is in excellent agreement with the predicted value. 
The same result  is retrieved  from the absorption curves depicted in 
Fig. \ref{fig:Absorbance}.
For making  explicit the dependence of the 
frequency of excitation of SPPs on the Fermi energy, we depict in Fig. \ref{fig:Absorbance} (bottom panel) the frequency corresponding to the maximum of absorption (in the SPP region)
as function of the Fermi energy and we fit the data points with a function of the form $C\sqrt{E_F}$
and found an excellent agreement.
  The large bump seen in the absorption curve at
low frequencies is due to 
the excitation of waveguide modes between graphene and 
the metal (see discussion ahead)
 and not to the excitation of SPPs: its
intensity increases with $E_F$ but does not disperses as $E_F$ varies.
This is confirmed making a density plot of the fields in that frequency range: the intensity of the fields is (almost) zero everywhere, except in the gap region between graphene and the metal.

\section{ SPP exciation efficiency at the graphene sheet\label{sec:Excitation-of-a}}

In this section, we determine the spatial probability density, $|\alpha_{\pm}(x)|^{2}$, for exciting SPP's. The quantities $|\alpha_{+}(x)|^{2}$ and  $|\alpha_{-}(x)|^{2}$ are the spatial probability density of a SPP propagating to the right and to the left of the slit, respectively. In order to compute it, we must define the magnetic and electric fields on the graphene sheet. Graphene acts as an open waveguide, therefore, supporting both SPP and radiative modes, allowing us to write the composite fields as a superposition of the two types of modes
\cite{Lalanne2005-PRL}
\begin{equation}
H_{y}(x,z)=[\alpha^{+}(x)+\alpha^{-}(x)]H_{y,spp}+\int d\rho\,a(x,\rho)H_{y,rad},\label{M-field-2-1}
\end{equation}
and
\begin{equation}
E_{z}(x,z)=[\alpha^{+}(x)-\alpha^{-}(x)]E_{z,spp}+\int d\rho\,a(x,\rho)E_{z,rad},\label{eq:E-fields-2-1}
\end{equation}
where $H_{y,spp}$ and $E_{z,spp}$ represent the SPP modes and $H_{y,rad}$
and $E_{z,rad}$ the radiative modes. The modes are orthogonal \cite{chaves2018scattering2} as long as we neglect dissipation:
\begin{equation}
\int dzH_{y,spp}E_{z,rad}=\int dzH_{y,rad}E_{z,spp}=0.\label{eq:ortho}
\end{equation}
Therefore, in order to compute the probability density, we must obtain the
SPP fields. In addition, it will be only necessary to calculate $\alpha_{+}(x)$,
since, by symmetry, $\alpha_{+}(x)=\alpha_{-}(-x)$.

\subsection{The probability density $|\alpha_{+}(x)|^{2}$ of exciting a SPP\label{subsec:The-probability-}}

On physical grounds, we can assume that, for $x>w$, we have $\alpha_{-}(x)\approx0$ \cite{lalanne2006_approximate},
and using the orthogonality relations
the result for $\alpha_{+}(x)$ follows:
\begin{equation}
\alpha^{+}(x)=\frac{\int dzH_{y}E_{z,spp}+\int dzE_{z}H_{y,spp}}{2\int dzH_{y,spp}E_{z,spp}}.\label{alpha-equation}
\end{equation}

The fields $H_{y}$ and $E_{z}$ have already been determined  and the SPP fields are obtained in Ref. \cite{chaves2018scattering} (see also supplementary material).
Hence, the $z- $dependence of the magnetic SPP field reads

\begin{equation}
H_{y,spp}(x,z)=\bar{H}_{y,spp}h(z)e^{iqx},
\end{equation}
while the nonzero components of the electric field are:
\begin{equation}
E_{x,spp}(x,z)=-i\frac{\mu_{0}c^{2}}{\omega\epsilon_{\alpha}}\bar{H}_{y,spp}h^{\prime}(z)e^{iqx},
\end{equation}
and
\begin{equation}
E_{z,spp}(x,z)=-\frac{q\mu_{0}c^{2}}{\omega\epsilon_{\alpha}}\bar{H}_{y,spp}h(z)e^{iqx},
\end{equation}
where
\begin{equation}
h(z)=\begin{cases}
u\cosh(\kappa_{S}z), & 0<z<d,\\
ve^{-\kappa_{1}(z-d)}, & z>d.
\end{cases}
\end{equation}
Inserting the definitions of the fields into (\ref{alpha-equation}),
we get, for the numerator
\begin{align}
 & \int_{-\infty}^{\infty}dzH_{y,spp}E_{z}+\int_{-\infty}^{\infty}dzE_{z,spp}H_{y}=\nonumber \\
 & =-\frac{\mu_{0}c^{2}}{\omega}\bar{H}_{y,spp}\int_{-\infty}^{\infty}dke^{i(k+q)x}(k+q)\nonumber \\
 & \times\left\{ \frac{e^{-i\zeta_{S,k}d}u}{\epsilon_{S}\left(\kappa_{S}^{2}+\zeta_{S}^{2}\right)}\left[i(A-B)e^{i\zeta_{S,k}d}\zeta_{S,k}\right.\right.\nonumber \\
 & +i\left(B-Ae^{2i\zeta_{S,k}d}\right)\zeta_{S,k}\cosh(\kappa_{S}d)\nonumber \\
 & \left.\left.+\left(B+Ae^{2i\zeta_{S,k}d}\right)\kappa_{S}\sinh(\kappa_{S}d)\right]+\frac{v}{\epsilon_{1}}\frac{e^{i\zeta_{1,k}d}}{\kappa_{1}-i\zeta_{1}}t_{k}\right\} ,
\end{align}
and, for the denominator,
\begin{align}
 & 2\int_{-\infty}^{\infty}dzH_{y,spp}E_{z,spp}=-\frac{q\mu_{0}c^{2}|\bar{H}_{y,spp}|^{2}e^{2iqx}}{\omega}\nonumber \\
 & \qquad\times\left[\frac{u^{2}}{\epsilon_{S}}\left(\frac{2\kappa_{S}d+\sinh(2\kappa_{S}d)}{2\kappa_{S}}\right)+\frac{v^{2}}{\epsilon_{1}\kappa_{1}}\right],
\end{align}
where $u^{2}$ and $v^{2}$ are given by \cite{chaves2018scattering}:
\begin{align}
u^{2} & =\frac{8w\epsilon_{S}\omega}{q\sqrt{\epsilon_{2}}c}\nonumber \\
 & \times\left[2d+\frac{\sinh(2\kappa_{S}d)}{\kappa_{S}}+\frac{2\epsilon_{S}\epsilon_{1}\epsilon_{0}^{2}\omega^{2}\cosh^{2}(\kappa_{S}d)}{\kappa_{1}(i\kappa_{1}\sigma+\epsilon_{0}\epsilon_{1}\omega)^{2}}\right]^{-1},
\end{align}
\begin{equation}
v^{2}=\left[\frac{\epsilon_{1}\kappa_{S}}{\epsilon_{S}\kappa_{1}}\sinh(\kappa_{S}z)\right]^{2}u^{2}.
\end{equation}

\begin{figure}[h]
\begin{centering}
\includegraphics[width=7cm]{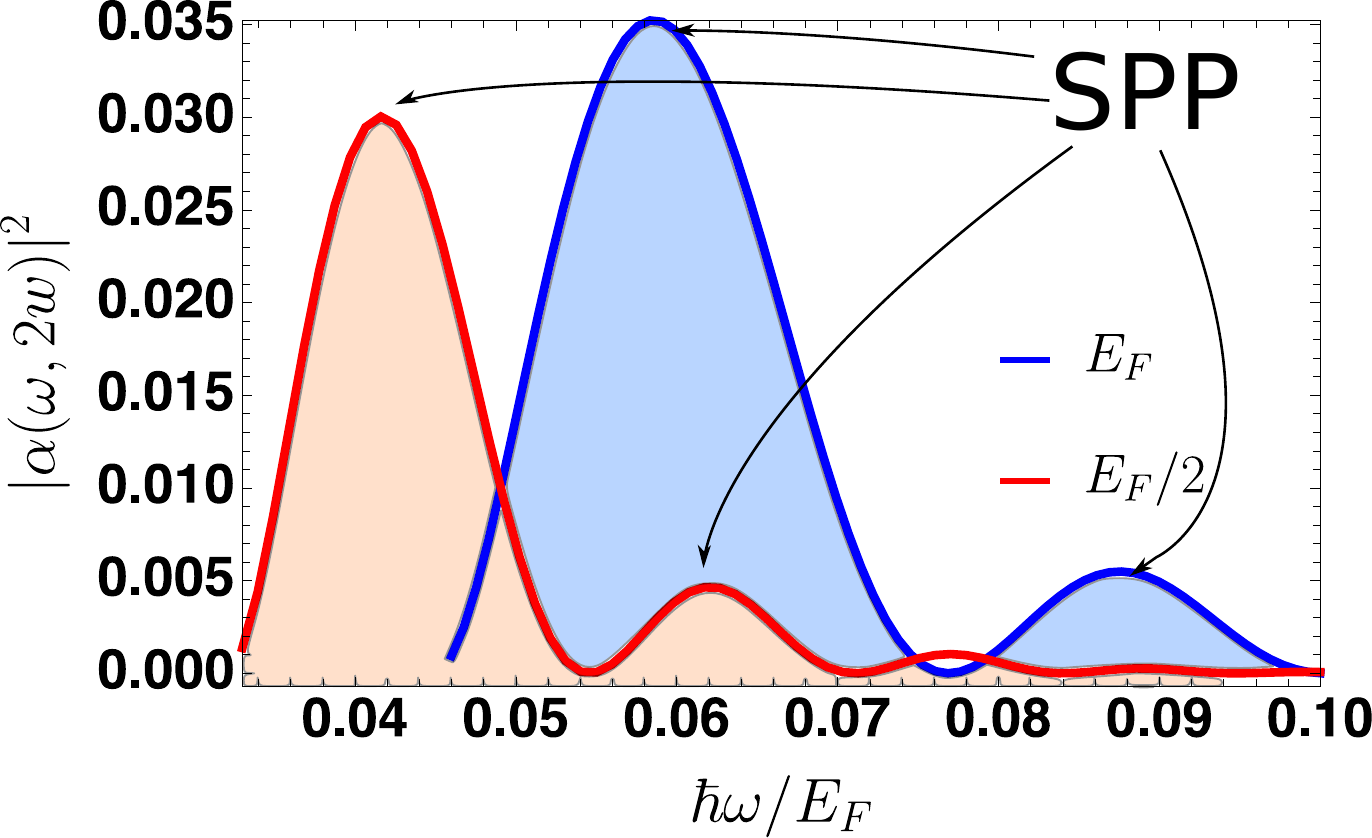}
\par\end{centering}
\caption{The probability $|\alpha(\omega,x)|^{2}$ as a function of the frequency
per Fermi energy, with $x=2w$, $E_{F}=1\,$eV, $\epsilon_{1}=1$,
$\epsilon_{S}=1.5$, $\epsilon_{2}=1$, $w=2\,\mu \rm{m}$, $\Gamma= 10^{-3}\,$eV, and $d=0.3\,\mu \rm{m}$.
For each curve we observe two peaks, where the ratio between the
first peaks of the blue and red lines are approximately equal to
$\sqrt{2}$. This also happens 
for the second peaks of the curves. This means that the frequencies
of such peaks are associated with the excitation of SPP's in graphene, as
discussed in the text.\label{fig:The-probability-}}
\end{figure}

In Fig. \ref{fig:The-probability-}, we plot the spatial probability density 
 $|\alpha(x)|^{2}$
in terms of the frequency in units of the Fermi energy, taking two different
values for this latter quantity. There are two well visible peaks (for each curve) in the region
$0.05<\hbar\omega/E_{F}<0.1$, with the amplitude of these peaks 
being approximately constant for different  Fermi energies. The ratio between the first peak of $\vert \alpha(E_{F}/2)\vert^2$
at $\omega/E_{F}\approx0.0416$ (red curve) and $\vert\alpha(E_{F})\vert^2$ at $\omega/E_{F}\approx0.0583$ (blue curve)
is approximately equal to $\sqrt{2}$.  This is in agreement with previous discussion in connection with Eq.  (\ref{eq_ratio}) and shows that these peaks in $|\alpha(\omega,x)|^{2}$ are associated with the excitation of surface-plasmons in graphene.
We obtain the same result if we take the ratio between the second (less intense) peaks
of the $E_{F}$ and $E_{F}/2$ curves, which in this case  Eq.  (\ref{eq_ratio}) also predicts a value of $\sqrt{2}$.

Figure \ref{fig:The-probability-} also allows to 
extract information about the wave-number of the 
SPP. For a graphene sheet at large distance from the metal, the frequency of the SPP is proportional to the square-root of the wave-number $q$: 
$\omega\propto \sqrt{q}$,  whereas 
for close distances to the metal we have
$\omega\propto q$.
Let us fix our attention, for example, on the red curve of Fig. \ref{fig:The-probability-}. There are two well defined peaks. Assuming that  $q\propto n\pi/w$, where $n$
 is an integer, and labeling the frequency of the SPP by the number $n$, it is clear, for large
 distances, that$
 \frac{\omega_{n+1}}{\omega_{n}}=\sqrt{(n+1)/n .}
 $
 If we now read the frequency of the two maxima we find $\hbar\omega_1/E_F=0.087$ and  $\hbar\omega_2/E_F=0.058$. Taking the  ratio we find: $\omega_2/\omega_1=1.5\approx\sqrt 2$  in  approximate agreement with the previous frequency ratio.

\begin{figure}[h]
\centering{}\includegraphics[width=8cm]{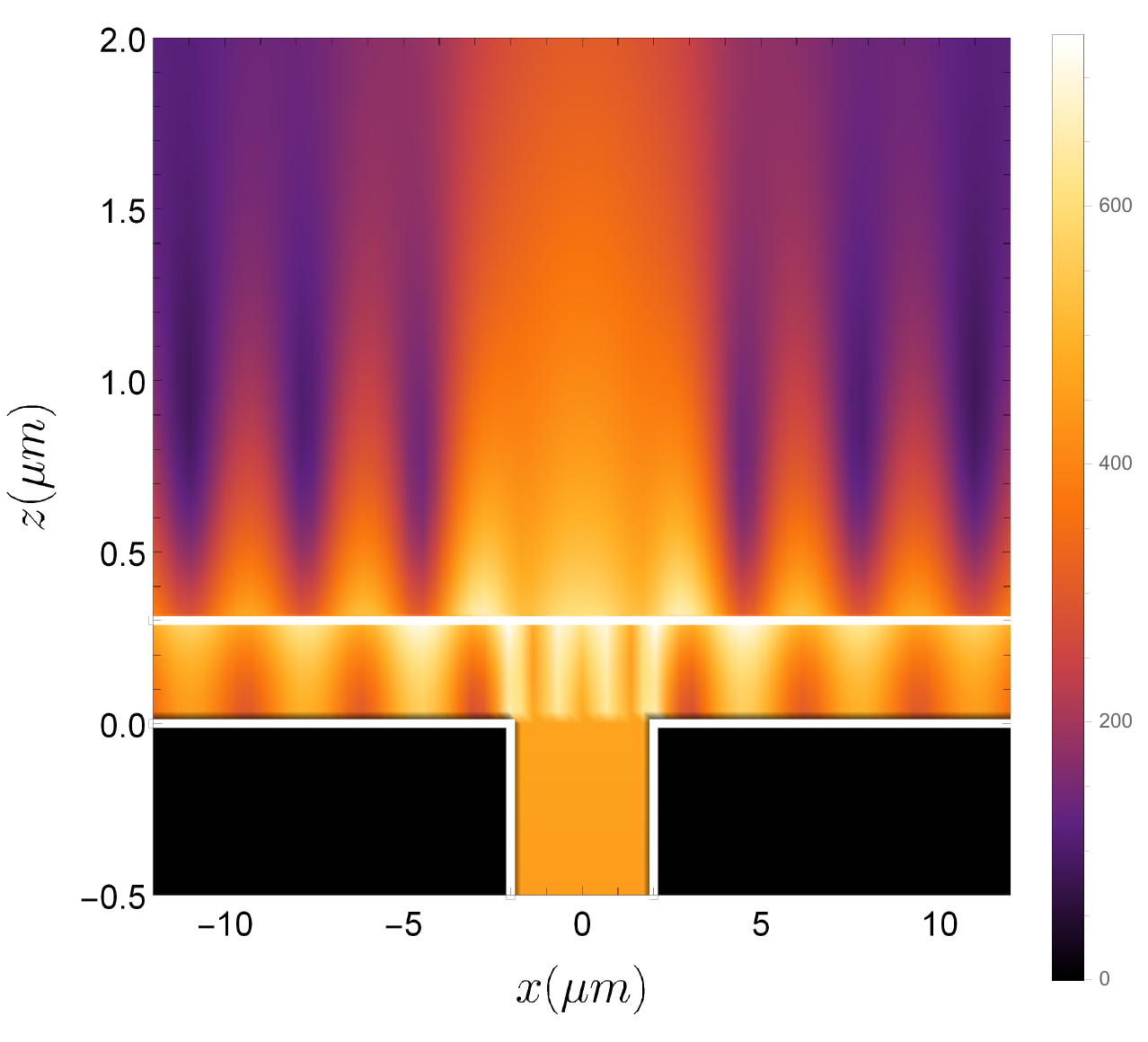}\caption{Density plot of the electric field, with $\hbar\omega/E_F=0.06$, $E_{F}=1\,$eV,
$\epsilon_{1}=1$, $\epsilon_{S}=1.5$, $\epsilon_{2}=1$, $w=2\,\mu \rm{m}$, $\Gamma=10^{-3}\,$eV, and $d=0.3\,\mu \rm{m}$.
The intense maxima and minima seen between graphene and the metallic screen is due to the SPPs forming in that region.The black regions
represent the metallic screen. The horizontal white line represents  the graphene sheet.
\label{fig:The-stream-density}}
\end{figure}

Choosing the frequency in Fig. \ref{fig:The-probability-} where
there is a peak, we should be able to extract the wavelength of the SPP  modes from the plots of the electric fields.
Therefore, we have chosen the frequency of the first plasmonic peak of $\vert \alpha(x)\vert^2$
for the $E_{F}$ curve in Fig. \ref{fig:The-probability-}, that is, the
peak located at $\hbar\omega/E_{F}=0.0669$. After extracting this
frequency, we  calculated the  density plot of the $z-$ and $x-$components
of the electric field in regions $S$ and $1$, given by Eqs. (\ref{eq:ESx}),
(\ref{eq:ESz}), (\ref{eq:E1x}), and (\ref{eq:E1z}). 
In Fig. \ref{fig:The-stream-density},
we clearly see, in the region between the metal and the graphene sheet, the existence of maxima and minima associated with the plasmon field. 

\begin{figure}[h]
\centering{}\includegraphics[width=7cm]{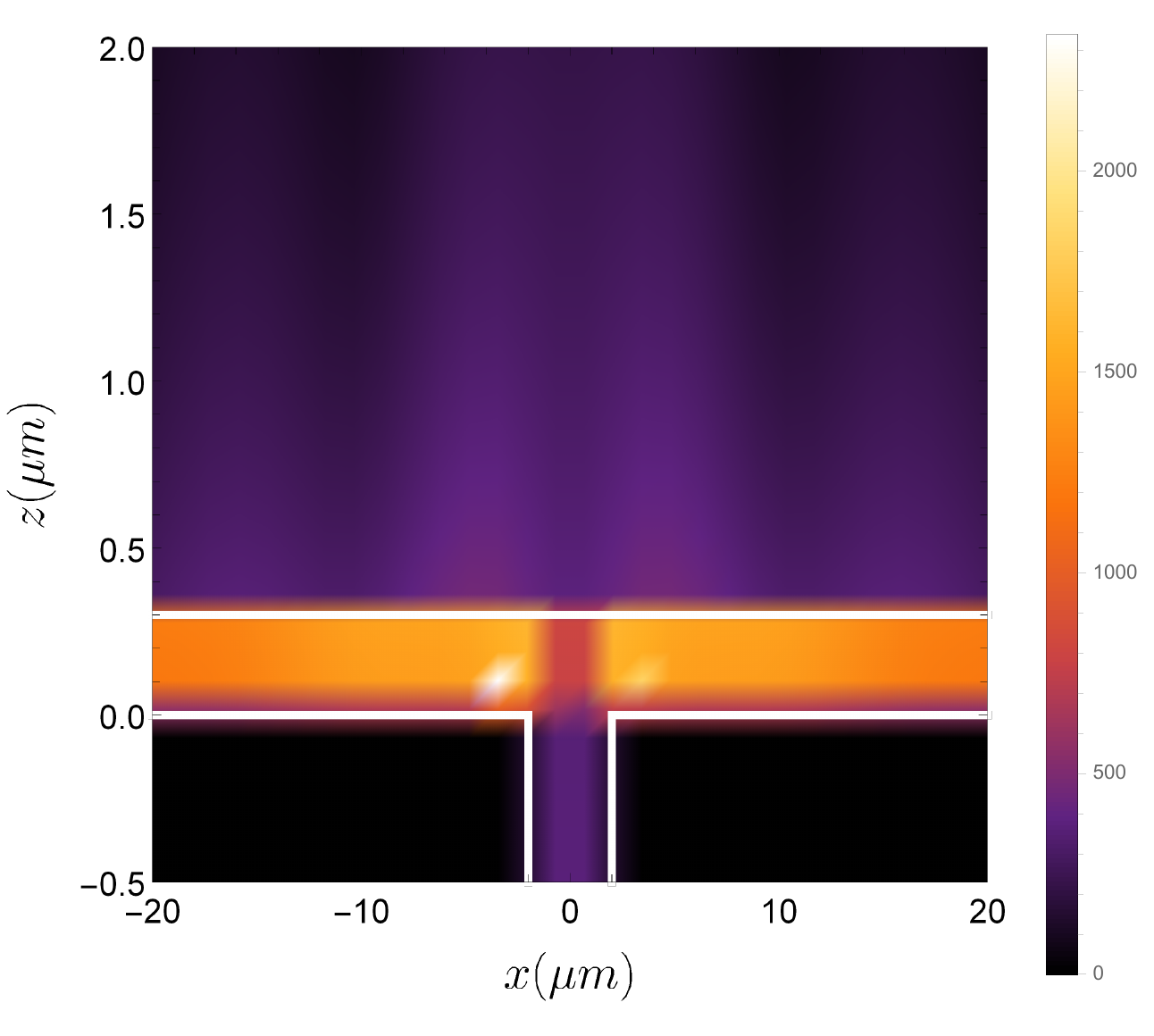}\caption{ 
Density plot of a waveguide mode. 
The largest intensity of the electric field occurs between graphene and the metallic screen.
The parameters are  $\hbar\omega/E_F=0.02$ (a frequency located in the yellow region of Fig. \ref{fig:Absorbance}), $E_{F}=1\,$eV, $\epsilon_{1}=1$, $\epsilon_{S}=1.5$,
$\epsilon_{2}=1$, $w=2\,\mu \rm{m}$,   $d=0.3\,\mu \rm{m}$, and $ \Gamma=10^{-3}\,$eV.
\label{fig:The-stream-density-eva}}
\end{figure}

For highlighting the waveguide modes in the low frequency region,
in Fig. \ref{fig:The-stream-density-eva} we
have considered the frequency $\hbar\omega/E_F=0.02$, which is located in the yellow region of Fig. \ref{fig:Absorbance}. It is clear 
from density plot that the field is very intense in the region between graphene and metallic screen and is very weak above the graphene sheet. This behavior
is characteristic of a waveguide mode.

\section{Final Remarks}

In this paper, we have considered a slit of width $w$
in a semi-infinite perfect metal screen, which is at a distance $d$
from a graphene sheet. The goal was to study  the excitation of SPP
modes on graphene.  We first defined the three regions of
interest, one inside the metal, the other between the graphene sheet
and the metal, and the last one below graphene. Then, we considered
only the fundamental mode of the incident field as the impinging one,
and, after applying the respective boundary conditions, we completely
determined the electric and magnetic fields for the three regions in a quasi-analytical manner. From this, we were able determine the reflection, transmission, and absorption in graphene.
From the knowledge of the fields, we could determine the spatial
probability density of exciting a SPP in graphene.
We have studied its dependence on the frequency of the incoming field and found peaks that  we were able to identify with the efficient excitation of SPP's.
The methods developed here can also be extended to gratings, with the fields represented by a Fourier series instead of a Fourier integral.

\acknowledgments{
D.C.P. was supported by Coordenação de Aperfeiçoamento de Pessoal
de Nível Superior (CAPES/Brazil), through Programa de Doutorado Sanduíche
no Exterior (PDSE) -- Process 88881.187657/2018-01, and also thanks
the hospitality of the Centro de Física, Universidade do Minho, Braga
-- Portugal. N.M.R.P. acknowledges support from the European Commission
through the project \textquotedblleft Graphene Driven Revolutions
in ICT and Beyond\textquotedblright{} (Ref. No. 785219), FEDER, and
the Portuguese Foundation for Science and Technology (FCT) through
project POCI-01-0145-FEDER028114, and in the framework of the Strategic
Financing UID/FIS/04650/2.
}

\bibliographystyle{epl2-author/eplbib}

\end{document}